# Multiscale phonon blocking in Si phononic crystal nanostructures

M. Nomura[1,2,*], Y. Kage[1], J. Nakagawa[1], T. Hori[3], J. Maire[4], J. Shiomi[3], D. Moser[5], and O. Paul[5]

[1]Institute of Industrial Science, The University of Tokyo, Tokyo 153-8505, Japan
[2]Institute for Nano Quantum Information Electronics, The University of Tokyo, Tokyo 153-8505, Japan
[3]Department of Mechanical Engineering, The University of Tokyo, Tokyo, 113-8656, Japan
[4]LIMMS/CNRS-IIS, The University of Tokyo, Tokyo 153-8505 Japan
[5]Department of Microsystems Engineering (IMTEK), University of Freiburg, Freiburg, 79110, Germany
*nomura[at]iis.u-tokyo.ac.jp

In-plane thermal conduction and phonon transport in both single-crystalline and polycrystalline Si two-dimensional phononic crystal (PnC) nanostructures were investigated at room temperature. The impact of phononic patterning on thermal conductivity was larger in polycrystalline Si PnCs than in single-crystalline Si PnCs. The difference in the impact is attributed to the difference in the thermal phonon mean free path (MFP) distribution induced by grain boundary scattering in the two materials. Grain size analysis and numerical simulation using the Monte Carlo technique indicate that grain boundaries and phononic patterning are efficient phonon scattering mechanisms for different MFP length scales. This multiscale phonon blocking structure covers a large part of the broad distribution of thermal phonon MFPs and thus efficiently reduces thermal conduction.

## I. INTRODUCTION

Nanoscale heat transport has recently been investigated because of its importance both in fundamental physics and practical applications[1,2]. Characteristic phonon transport phenomena have been reported in a variety of systems such as semiconductor super lattices[3,4], membranes containing nanoparticles[5], porous nanostructures[6-8], nanowires[9], and phononic crystal (PnC) structures[10-13]. The phonon mean free path (MFP) is the most important physical parameter for determining the thermophysical property of a system. The thermal conductivity of a crystal in a bulk material is reduced when the characteristic length of the system becomes comparable to or shorter than the phonon MFP in the bulk material. In such a system, the thermal property no longer follows the Fourier law but instead is described by semiballistic phonon transport. Phonon transport physics is especially interesting in PnC nanostructures because the wave nature of phonons appears[14] where coherence is preserved, and together with the incoherent phonon scattering process[15], determines phonon transport.

Thermoelectrics is an important application, in which phonon transport nanoengineering plays an active role. The figure of merit $ZT$ is given by $S^2\sigma T/\kappa$, where $T$ is temperature, $S$ is the Seebeck coefficient, $\sigma$ is electrical conductivity, and $\kappa$ is thermal conductivity. The concept of "phonon-glass and electron-crystal" proposed by Slack yields the following strategy for increasing $ZT$: reduce thermal conduction while maintaining electrical conduction by taking advantage of different phonon and electron MFPs[16]. Many groups have demonstrated very low values of thermal conductivity by increasing phonon scattering in a variety of nanostructures[1]. However, for this purpose, it is important to be aware that thermal phonons are distributed over a broad range of frequencies[17,18], and thus there is a multiscale distribution of thermal phonon MFPs. Therefore, a phonon blocking mechanism that covers the whole MFP range is required in order to reduce thermal conductivity efficiently. Biswas *et al.* demonstrated the validity of this idea using an all-scale hierarchical architecture in which three different scattering mechanisms blocked phonon transport and covered different thermal phonon MFP ranges from the atomic scale to the mesoscale by using PbTe (Ref. 19).

Si is considered one of the most promising candidates of a high performance thermoelectric material with a low environmental load. Therefore, it is important to investigate the validity of the strategy for Si and to obtain quantitative information for structural design.

In this study, we investigated the use of all-scale hierarchical architecture in Si for efficient reduction of thermal conduction. We fabricated two-dimensional (2D) PnC nanostructures using both single-crystalline and polycrystalline Si membranes and measured thermal conductivity at room temperature. In the polycrystalline Si PnC nanostructures, grain boundary scattering blocked thermal phonons with relatively short MFPs, and phononic patterning blocked phonons with longer MFPs. We discuss the validity of multiscale phonon blocking in Si, using grain size analysis data and numerical simulation results generated via a Monte Carlo technique.

## II. EXPERIMENTAL DETAILS

The PnC nanostructures were fabricated with either single-crystalline or polycrystalline Si membranes. For single-crystalline Si samples, we used a commercially available (100) nominally boron-doped silicon-on-insulator wafer with a 145 nm-thick upper Si layer and a 1 μm-thick $SiO_2$ buried oxide layer. For polycrystalline Si samples, the top polycrystalline Si layer was grown on a 1.5 μm-thick $SiO_2$ layer on a (100) Si wafer by low-pressure chemical vapor deposition at 625 °C. The layer was undoped and was 143 nm thick, as measured by ellipsometry. The PnC nanostructures were formed via electron beam (EB) lithography using a reactive ion etching/inductively coupled plasma system, with $SF_6/O_2$ gas as the etchant. The oxide layer under the Si layer was removed with hydrofluoric acid in order to form the air-bridge structures. The dimensions of the PnC nanostructures were measured by scanning electron microscopy (SEM).

Figure 1(a) shows an SEM image of the whole air-bridge structure with the 2D PnC structures. In the center of the air-bridge structure, a 125 nm-thick Al layer was deposited to form a 4 × 4 μm² pad on the Si layer to enable thermoreflectance measurements. The central Si island with the Al pad atop was then supported by two PnC structures. The width and the length of the structure were 5 μm and 25 μm, respectively. The 2D PnC structures were formed using circular air holes periodically aligned as a square lattice, as shown in Fig. 1(b). PnC structures with a variety of radii were fabricated with a fixed period $a = 300$ nm.

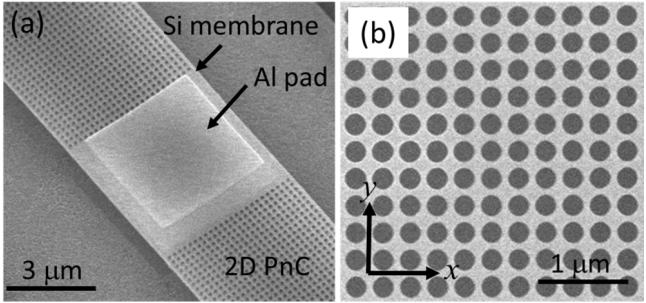

FIG. 1. SEM images of (a) the whole air-bridged Si PnC nanostructure and (b) a 2D PnC nanostructure with circular air holes aligned as a square lattice ($a = 300$ nm and $r = 115$ nm).

To measure in-plane thermal conductivities of the PnC nanostructures, a method using micro time-domain thermoreflectance (μ-TDTR) was developed, which has very high throughput as compared to electrical methods. The principle of this technique is the same as that for TDTR[20]; however, it can be applied to micrometer-sized systems. The Al pad was heated by a quasicontinuous laser beam (wavelength $\lambda = 642$ nm) for 500 ns, and the temporal evolution of the temperature of the Al pad (the TDTR signal) was monitored by a continuous-wave laser beam ($\lambda = 785$ nm). Both beams were collinearly focused on the Al pad, as shown in Fig. 2(a), by using a microscope objective lens with a numerical aperture of 0.65. The beam spot on the pad was approximately 700 nm in diameter. The power of the pump pulse was set so the temperature increase was less than 5 K in the Al pad, and it was verified that the pump power did not change the thermal conductivity in the range of temperature increase. We calculated the radiation loss by Stefan-Boltzmann law and verified that the radiation loss is negligible under this experimental condition. All the measurements were performed at room temperature and in a vacuum chamber to eliminate heat-convection loss. The PnC nanostructures that supported the central Si island were the only heat dissipation channels. This well-defined system made it possible to obtain a highly reliable value of the thermal conductivity by analyzing the temporal evolution of the TDTR signal. The temperature evolution in whole structure was simulated via the finite element method (FEM) by using COMSOL Multiphysics®. Figure 2(b) shows the simulated temperature distribution, 15 μs after pulse heating, for a PnC structure with $r = 115$ nm. All the structural parameters were obtained by SEM measurement, and the only fitting parameter for the measured TDTR signal was thermal conductivity. The value of thermal conductivity was obtained via the least square method. The details of the analysis can be found in our previous publication[21].

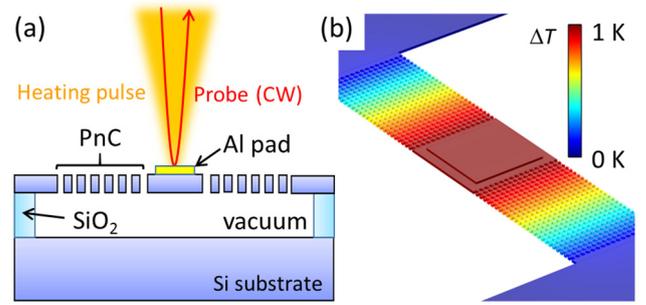

FIG. 2. (Color online) (a) Schematic of the μ-TDTR technique and cross section of the air-bridge structure. (b) Simulated temperature distribution at 15 μs after pulse heating a PnC structure with $a = 300$ nm and $r = 115$ nm.

### III. EXPERIMENTAL RESULTS

For both single-crystalline and polycrystalline samples, the thermal conductivity of the PnC nanostructure was measured by μ-TDTR at room temperature. Figure 3 shows TDTR signals for single-crystalline PnCs with $r = 73$ nm, 115 nm, and 126 nm, as well as for an unpatterned membrane. The dots and the solid lines represent experimental and simulated TDTR traces, respectively. The increase in the TDTR signal was caused by the temperature increase in the Al pad from the heating pulse, which was applied from time $t = 0$ to $t = 500$ ns. The heat was only dissipated through the PnC nanostructures, and the temperature of the Al gradually decreased on a timescale on the order of tens of microseconds. The best fitting curve for each PnC with $r = 73$, 115, and 126 nm, as well as for the unpatterned membrane, is given by thermal conductivities of 41 Wm$^{-1}$ K$^{-1}$, 35 Wm$^{-1}$ K$^{-1}$, 29 Wm$^{-1}$ K$^{-1}$, and 75 Wm$^{-1}$ K$^{-1}$, respectively.

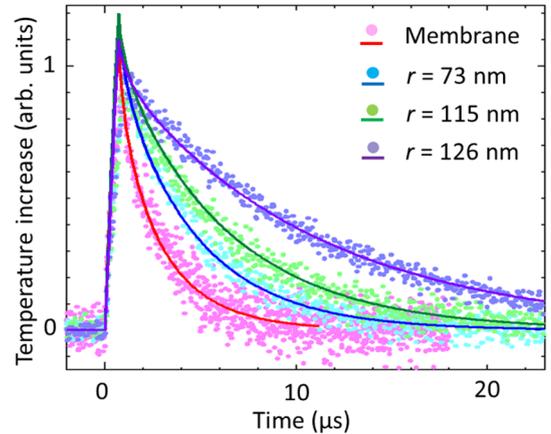

FIG. 3. (Color online) TDTR signals for the unpatterned membrane and the single-crystalline Si PnC nanostructures ($r = 73$ nm, 115 nm, and 126 nm). The dots and solid lines represent experimental and simulated TDTR signals, respectively.

The thermal conductivity of the unpatterned membrane is similar to the reported value for a single-crystalline Si membrane of the same thickness[22], thus confirming the reliability of the measurement and analysis.

Thermal conductivity was also measured in the polycrystalline PnC structures. Figure 4 summarizes the thermal conductivities obtained by FEM simulation for both types of crystalline Si PnC nanostructure, showing a decrease in thermal conductivity as the radius of the air holes decreases. This indicates that the characteristic lengths of the PnC nanostructures were comparable to the MFPs of thermal phonons and that the system was in the semiballistic phonon transport regime. We found that forming air holes, even when the value of *r*/*a* was small, dramatically reduced thermal conductivity. This result indicates that the side walls of the air holes, which are perpendicular to the direction of in-plane phonon transport, backscatter phonons and largely reduce thermal conductivity[13]. For single-crystalline Si, the thermal conductivity showed a steep reduction above *r* = 120 nm. In this regime, where the neck size, $a - 2r$, was smaller than the thickness (145 nm), the necking effect[7] became more effective and dramatically reduced thermal conductivity. This tendency had also been observed in our previous measurements for Si nanowires[21].

For polycrystalline PnC structures, the measured thermal conductivity for the membrane was 11 Wm$^{-1}$ K$^{-1}$, and for PnCs with *r* = 82 nm, 98 nm, and 123 nm, the thermal conductivity was 5.0 Wm$^{-1}$ K$^{-1}$, 4.7 Wm$^{-1}$ K$^{-1}$, and 4.3 Wm$^{-1}$ K$^{-1}$, respectively. The radius dependence was similar to that of the single-crystalline samples, but the impact of the phononic patterning was different because of the different thermal phonon MFP distributions. The magnitudes of the reduction in thermal conductivity for single-crystalline and polycrystalline Si PnCs can be compared by considering the data around *r* = 100 nm, where the necking effect is moderate. By interpolating the experimental plots, the thermal conductivities at *r* = 100 nm plots were estimated as 38 Wm$^{-1}$ K$^{-1}$ and 4.66 Wm$^{-1}$ K$^{-1}$ for the single-crystalline and the polycrystalline samples, respectively. The magnitudes of the reductions in membrane thermal conductivity were 49% and 58%, respectively. This indicates that the impact of the phononic patterning is larger for polycrystalline Si PnC nanostructures than for single-crystalline nanostructures.

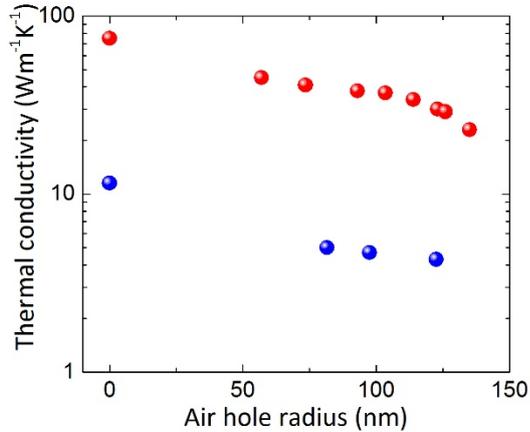

FIG. 4. (Color online) Measured thermal conductivities for a variety of radii in single- (red circles) and polycrystalline (blue circles) Si PnC nanostructures.

The difference in the magnitudes of the reduction in thermal conductivity can be explained by examining nano- and mesoscopic phonon transport. Figures 5(a) and 5(b) show top and cross-sectional images taken by a transmission electron microscope (TEM). The in-plane grain shape is random, and fan-like growth can be observed in Fig. 5(b). The grain size analysis on the surface is summarized in the histogram shown in Fig. 5(c). The grain size was determined by approximating the random grain shape, using a circular shape with the same area. The grain sizes were mostly distributed around the 20 nm mark and with sizes on the order of a few nanometers to up to 60 nm. However, due to the conical shape of the grains, the histogram is shifted toward smaller sizes. The TEM analysis indicates that phonons which have MFPs that are similar in size to the grains in single-crystalline Si (1–60 nm) are scattered more frequently by grain boundaries in polycrystalline Si than in single-crystalline Si. Therefore, thermal phonon MFPs are distributed over a larger MFP regime in polycrystalline Si than in single-crystalline Si. However, the characteristic lengths of the PnC structures, which are approximately the size of the neck (>50 nm), shorten the phonon MFPs to this length. The latter mechanism of phonon blocking is more efficient than the former for phonons with MFPs longer than the neck size. Therefore, the crystal grain and phononic patterning cover different ranges of phonon MFPs, which results in efficient phonon transport blocking. Thus, this multiscale phonon blocking is a Si analog of the all-scale hierarchical architecture demonstrated by Biswas *et al.* using PbTe (Ref. 19).

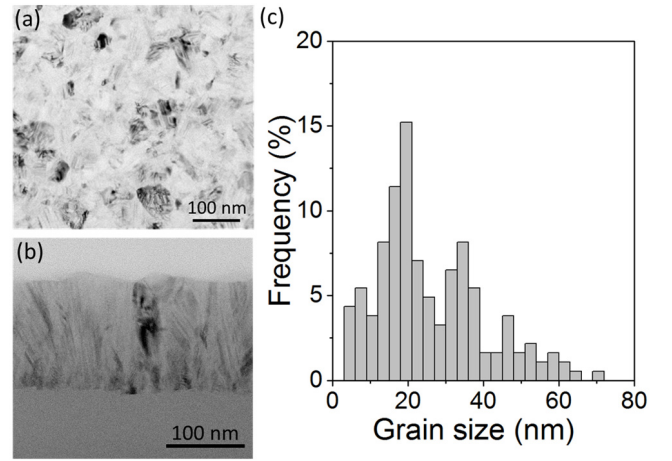

FIG. 5. TEM images of (a) top and (b) cross-sectional views for the poly-Si wafer. (c) Histogram of the grain size measured on the surface. The grain size at the bottom part is much smaller due to the conical shape of the grains.

## IV. NUMERICAL SIMULATION

To quantify the multiscale phonon blocking, thermal conductivity accumulation functions for bulk Si and for single-crystalline Si PnC nanostructures were calculated. In nanostructures, MFPs of phonons are shortened due to phonon boundary scattering. By defining the effective MFP as $\Lambda_{\text{eff}}$, nanostructure thermal conductivity in terms of Boltzmann transport can be written as

$$\kappa = \frac{1}{3}\sum_{\mathbf{q},s} c_{\mathbf{q},s} v_{\mathbf{q},s} \Lambda_{\text{eff},\mathbf{q},s} , \qquad (1)$$

where *c* and *v* are the specific heat and group velocity, which depend on phonon wave vector **q** and branch *s*.

The calculation of $\Lambda_{\text{eff}}$ was realized using a Monte Carlo ray tracing simulation. The simulation reproduces the single-crystalline Si PnC nanostructures in the experiment [Fig. 1(b)] by taking a square unit cell (side *a* = 300 nm) with an air hole in the center. The thickness of the unit cell was 150 nm. *N* unit cells were arranged in a row in the direction parallel to heat flux (*x* direction), and periodic boundary condition was imposed in the other in-plane direction (*y* direction).

The method calculates the phonon transmission probability (*τ*) by emitting a phonon from one side of the nanostructure (*x* = 0) with an incident polar (*θ*) and azimuthal (*φ*) angles, and statistically evaluating the probability of phonons to reach the other side ($x = Na$). When the phonon arrives at the out-of-plane boundaries or the surface of the air holes, it is scattered (reflected) fully diffusely. In addition, the phonon is scattered by phonon-phonon scattering i.e. it changes the direction randomly after traveling a distance $-\Lambda \ln R$ from the location of the previous scattering. Here, $\Lambda$ is the MFP of the phonon in bulk single crystal and $R$ is a random number between 0 and 1. As the value of $\tau$ exhibits the size effect when the phonon MFP is longer than the length of the simulated system $L = Na$, $L$ was chosen to be sufficiently large to obtain a converged value.

The transmission probability $\tau$ was calculated for a range of $\Lambda$. For each value of $\Lambda$, simulations scan over the polar angles $\theta$ [0, $\pi/2$] discretized into regular intervals ($\pi/180$), while the azimuthal angle $\varphi$ and the $y$-$z$ coordinates of emission are randomly chosen with uniform distributions. The simulations were performed 96,040 times for each value of $\theta$, and the values of $\tau$ were averaged. $\Lambda_{\text{eff}}$ can then be obtained based on kinetic theory and Landauer's formula as,

$$\Lambda_{\text{eff}} = \frac{3L}{2M} \int_0^{\pi/2} \tau(\theta, \Lambda) \cos\theta \sin\theta \, d\theta \,, \qquad (2)$$

where $M$ is a modification factor of the structure accounting for the porosity of PnC structures[23] calculated by COMSOL Multiphysics®.

Equation (2) shows that the dependence of $\Lambda_{\text{eff}}$ on **q** and $s$ ($\Lambda_{\text{eff},\mathbf{q},s}$) can be obtained by the dependence of $\Lambda$ on **q** and $s$ ($\Lambda_{\mathbf{q},s}$). $\Lambda_{\mathbf{q},s}$ here was obtained by lattice dynamics calculation of bulk Si at room temperature using the interatomic force constants from first principles[24]. Finally thermal conductivity of the single-crystalline Si PnC nanostructures was calculated by substituting $\Lambda_{\text{eff},\mathbf{q},s}$ into Eq. (1). We also calculated thermal conductivity of the Si membrane without the holes (i.e. $r = 0$) and confirmed agreement with the analytical solution of Boltzmann transport equation[25, 26].

Figure 6(a) shows the thermal conductivities (squares) calculated using the data for single-crystalline Si structures. The radius dependence of the calculated thermal conductivities reproduces the same tendency and nearly the same values. Figure 6(b) shows the calculated thermal conductivity accumulation functions[27] for bulk Si and for single-crystalline Si PnC nanostructures with $r$ = 30 nm, 90 nm, and 135 nm. The vertical coordinate of any point on the function represents the thermal conductivity of a phonon whose MFP is less than the horizontal coordinate of that point. The surface scattering by the PnCs shortens the $\Lambda_{\text{eff}}$ of phonons which have MFPs that are longer than the neck size in the unpatterned membrane and shifts the distribution functions to the shorter side. Note that the thermal conductivity accumulation is almost saturated around sizes that are similar or a few times longer than the neck size.

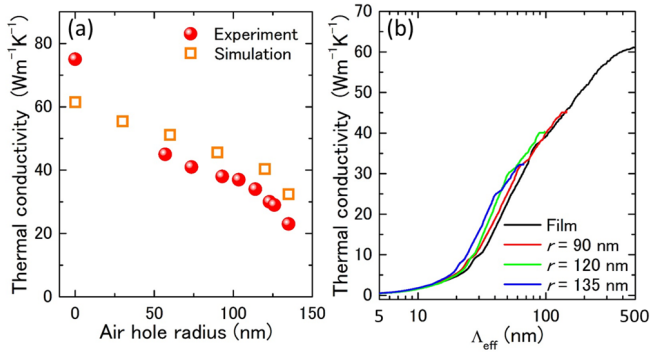

FIG. 6. (Color online) (a) Calculated cumulative thermal conductivities for bulk Si and for single-crystalline Si PnC nanostructures with various radii. (b) Radius dependence of the measured (red circles) and simulated (orange squares) thermal conductivities.

## V. DISCUSSION

The TEM analysis and the numerical simulations indicate that the grain size distribution in polycrystalline PnC nanostructures is between a few nanometers and 60 nm, as shown in Fig. 5(c), and that phononic patterning shortens phonon MFPs to 70–130 nm, depending on $r$, as shown in Fig. 6(a). These two different scattering mechanisms, boundary scattering and the surface scattering by the PnCs, cover almost the entire range of thermal phonon MFPs in Si. Therefore, this work demonstrates that all-scale hierarchical architecture can be applied to Si by taking advantage of polycrystallization and air-hole patterning by EB lithography. These air holes need not be periodically aligned when they are used as surface scattering centers. In our case, using a periodicity of 300 nm and measuring at room temperature, phonon transport lay mostly in the incoherent regime, where the wave nature of phonons is moderate[15, 23]. However, there are a few reports that suggest that PnC patterning may play an important role in thermal conduction in PnC structures because of both reduced group velocity and phononic band gaps[11]. While the impact of the phononic patterning on the thermal conduction remains a contested topic, given a recent publication on PnC microstructures below 1 K (Ref. 12), both the particle and the wave nature of phonons should be taken into account in PnC nanostructures[15], especially at cryogenic temperatures. In such an ideal solid state phononic system, which is the acoustic wave analog of photonic crystals[28, 29], the use of PnCs should make it possible to control heat transport as well as phonon transport[30].

## VI. CONCLUSIONS

In-plane thermal conductivity and phonon transport in single-crystalline Si PnC nanostructures and polycrystalline Si PnC nanostructures were investigated at room temperature. The impact of the phononic patterning was larger for polycrystalline membranes than for single-crystalline membranes. TEM grain size analysis and MFP calculation via the Monte Carlo technique indicated that grain boundary scattering and surface scattering cover different phonon MFP ranges (1–60 nm and >70 nm, respectively). The multiscale phonon blocking described in this work is a Si analog of all-scale hierarchical architecture, which has been demonstrated to improve thermoelectric performance in PbTe. We conclude that it is important to block phonons over the whole range of thermal phonons MFPs by using appropriate materials and structural designs.

## ACKNOWLEDGMENTS

This work was supported by the Project for Developing Innovation Systems of MEXT, Japan, by KAKENHI (25709090), and by the Foundation for the Promotion of Industrial Science.